\def\s2{[S\,{\sc ii}]}
\def\dog{[O\,{\sc i}]}
\def\xo2{[O\,{\sc ii}]}
\def\fe2{[Fe\,{\sc ii}]}
\def\ca2{[Ca\,{\sc ii}]}

\documentstyle[11pt,paspconf]{article}

\begin{document}

\title{Emission-Line Probes of Circumnuclear Dust in AGNs}
\author{Joseph C. Shields$^1$, Richard W. Pogge$^2$, and Michael M. De 
Robertis$^3$}
\affil{$^1$Ohio University, Physics \& Astronomy Dept., Athens, OH 45701 USA}
\affil{$^2$Ohio State University, Astronomy Dept., Columbus, OH 43210 USA}
\affil{$^3$York University, Dept. Physics \& Astronomy, Toronto, ON M3J1P3
Canada}

\begin{abstract}
Emission lines that trace elements subject to strong depletion onto
grains provide a means for probing the dust content of AGNs.  Examples
include infrared \fe2 and optical \ca2 lines.  The excitation
mechanisms underlying the \fe2 lines remain controversial, resulting
in related disagreement over the gas-phase abundance of iron in
Seyfert narrow-line regions.  In this contribution we emphasize the
utility of the \ca2 features as a consistency test for claims of grain
destruction affecting the \fe2 lines.  A search for \ca2 emission in
NGC~1068 at the location of strongest \fe2 emission along the radio
jet yields a strong upper limit, but no detection.  This result suggests
that grains survive largely intact in a region that otherwise
shows strong evidence of shock processing.

\end{abstract}


\section{Introduction}

Dust in the circumnuclear environment can be important for modifying
our view of the narrow- and broad-line regions in AGNs.  Netzer \&
Laor (1993) have suggested further that dust in the emission-line
regions has a major role in establishing the NLR/BLR dichotomy.  Dust
is also clearly implicated in reprocessing a significant part of the
continuum luminosity generated on small scales, with resultant
emission appearing at infrared wavelengths.  A detailed knowledge
of dust properties in AGNs is thus desirable, but our understanding
of grains in these environments remains primitive.

Emission-line diagnostics provide one means of probing the dust
content of the NLR.  The basic strategy is to measure nebular line fluxes 
from elements subject to depletion from the gas phase; if these features
do not play a dominant role in cooling the plasma, their strength
can be expected to scale approximately in proportion to the gas-phase
abundance of the responsible element (see, e.g., Kingdon et al. 1995 and
references therein).  The degree of depletion can then be used to
infer the prevalence of grains incorporating this element.

\section{Infrared \fe2}

In the local interstellar medium, the gas phase fraction of iron is
$\sim 10^{-2}$ (Savage \& Sembach 1996), making this element a
potentially useful tracer of grains.  Iron is commonly detectable in
Seyfert galaxies through prominent narrow-line emission in \fe2 1.257
$\mu$m, 1.644 $\mu$m.  The understanding of these lines
and their observed strengths remains controversial, however, with
different researchers arguing for excitation via supernovae
(Greenhouse et al. 1991), other forms of shocks (e.g., Veilleux et
al. 1997), or photoionization (e.g., Mouri et al. 1993; Simpson et al.
1996).  The disagreement over excitation leads to a related divergence
of opinion on gas-phase abundance, with some authors claiming that
strong \fe2 emission implicates substantial grain destruction,
presumably in shocks.

While the energetics of Seyfert NLRs are probably not globally
dominated by shocks (e.g., Laor 1998), there are nonetheless good
reasons to take shocks seriously in conjunction with the observed
\fe2 emission.  First, there is a strong statistical correlation 
between \fe2 and radio continuum emission in Seyfert nuclei.
While at first glance this trend might be dismissed as an illustration
that more powerful galaxies radiate more at all wavelengths, Veilleux
et al. (1997) point out that the \fe2/radio correlation is considerably
stronger than correlations between \fe2 and X-ray luminosities, even
when hard X-rays (which are less susceptible to absorption) are considered.
If \fe2 emission is powered primarily by photoionization, a robust
correlation between the line and ionizing continuum luminosities might be
expected; the fact that the \fe2/radio correlation appears stronger
may implicate a significant role for mechanical heating of the nebula
(as traced by \fe2) by the radio plasma.

A second suggestion that shocks driven by radio plasma may be
important in generating \fe2 emission is seen in a {\sl spatial}
correlation between these two emission components in NGC~1068, one of
the best-studied sources in terms of spatial resolution.  Blietz et
al. (1994) have published a comparison of \fe2 1.644 $\mu$m emission
and 5 GHz radio maps, which shows that the \fe2 emission is
concentrated around the inner part of the extended radio structure, in
a region that might be expected to contain material that was shocked
in the earlier passage of the expanding jet.  While alternatives to
shocks (e.g. photoionization by radiation collimated parallel to the
jet) can be invoked to account for the nebular gas, the correspondence
between the radio and \fe2 morphologies is nonetheless suggestive.

Finally, Veilleux et al. (1997) report that Seyfert 2 nuclei exhibit
\fe2 1.257~$\mu$m emission line widths that are often broader than
widths for hydrogen recombination lines.  The larger widths for
the low ionization \fe2 lines may indicate that this emission
traces a kinematically disturbed component of gas in the NLR, consistent
with a connection to shocks.

\section{Optical \ca2}

An independent test of grain destruction in the NLR is provided by
measurements of \ca2 $\lambda\lambda$7291, 7324 (Ferland 1993).
Calcium tends to be highly depleted onto grains in the normal ISM,
with a typical gas-phase fraction of only $\sim 10^{-3}$.  Power-law
photoionization models appropriate for the NLR that assume solar
abundances in the gas phase predict strong emission in the \ca2 lines,
at a level comparable to or exceeding that in the more familiar red lines
of \dog, \s2, and \xo2 $\lambda\lambda$7320, 7331.  

The absence of significant \ca2 emission in the spectra of Seyfert
nuclei has been interpreted as evidence that the NLR hosts a largely
intact population of grains (Ferland 1993), with the corollary that
grain destruction in shocks is unimportant in these environments, at
least in some global sense.  An important consistency test comes from
observations of \ca2 in plasmas that are {\sl known} to be shocked.
Spectra of supernova remnants (e.g., Vancura et al. 1992) and
outflows from Herbig-Haro objects (e.g., Morse et al. 1996) show
significant emission in the \ca2 lines, validating the use of these
features as probes of grain destruction.

While shocks may be unimportant in the global energetics of the
narrow-line region, there are strong indications that shocks {\sl are}
important in localized portions of the NLR.  The evidence comes from
apparent shock structures that are either resolved in optical imaging
(e.g., Capetti et al. 1997) or inferred from kinematic correlations
with radio properties (Whittle et al. 1988).  These sites may
additionally play an important role in generating \fe2 emission seen
in the global spectra of the Seyfert NLRs.  A search for \ca2 emission
in these localized sites would provide a useful constraint on the
significance of shocks in these regions, while also directly probing
the degree of grain destruction.

\section{A Test Case: NGC~1068}

The prototypical Seyfert 2 galaxy NGC~1068 provides a useful testbed
for assessing the role of shocks within the NLR and the resulting
consequences for \fe2 and \ca2 emission.  The nucleus of this galaxy
features a radio jet with a well defined bow-shock structure (e.g.,
Wilson \& Ulvestad 1983), and correlated nebular emission (Capetti et
al. 1997).  The optical emission features exhibit line-splitting of up
to 1500 km s$^{-1}$, interpreted as a signature of gas motions driven
by the expanding synchrotron plasma (Axon et al. 1998).  As noted
previously (\S2), the correspondence between the radio and \fe2
emission structures provides circumstantial evidence favoring a role
for shocks in generating the \fe2 luminosity.

In order to test this scenario further, we have conducted a search for
\ca2 emission in the extended NLR in the environs of the detected
radio and \fe2 emission.  The data used for this purpose were acquired
in a one-hour exposure with the Canada-France-Hawaii 3.6-m telescope
in conjunction with the MOS/ARGUS integral field optical fiber
spectrometer.  The optical fibers project to a diameter of
$0.4^{\prime\prime}$, and the resulting data have a spectral
resolution of $\sim 5$\AA .

A spectrum corresponding to a $1.6^{\prime\prime} \times
1.1^{\prime\prime}$ aperture centered on the peak of the \fe2 emission
(northeast of the nucleus) is shown in Figure 1.  The expected
location of \ca2 $\lambda$7291 emission is indicated by the vertical
bar.  \ca2 $\lambda$7324, if present, is blended with \xo2, but the
$\lambda$7291 feature is expected to be the stronger line of the \ca2
doublet, with a theoretical intensity ratio of $I(\lambda7291)/I(\lambda(7324)
\approx 3$.

\begin{figure}
\vspace*{90mm}
\includegraphics{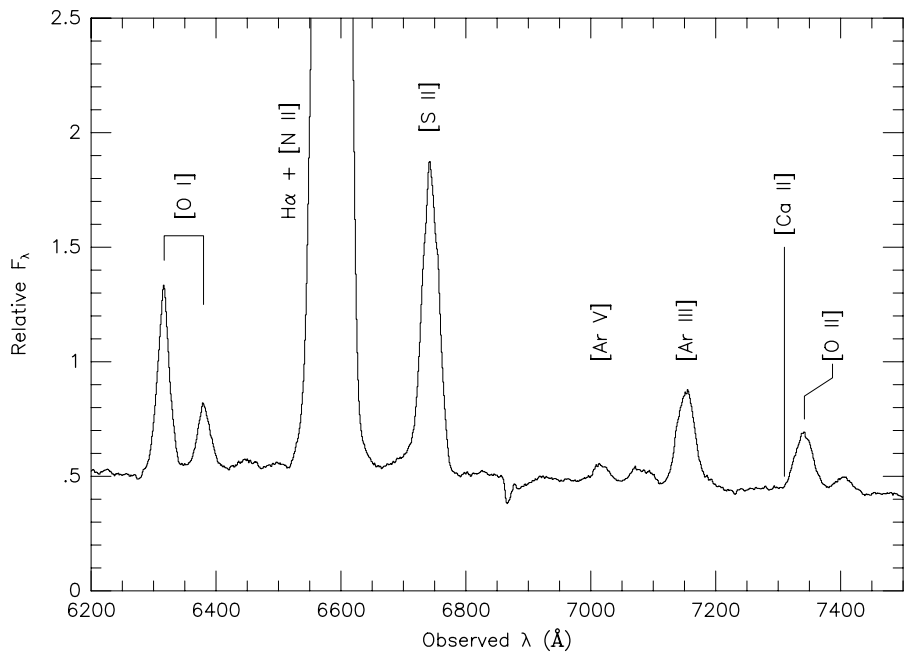}
\caption{Spectrum of NLR emission in NGC~1068, with the expected location of
\ca2 $\lambda$7291 indicated.}
\end{figure}
No \ca2 emission is apparent in the spectrum in Figure 1, and inspection
of other spectra from this data set for nearby regions in the extended
NLR of NGC~1068 similarly yield a negative result.  A useful way of
expressing this finding is as a 3-$\sigma$ upper limit of $I({\rm \ca2}
\lambda 7291)/I($[O\,{\sc i}]$\lambda 6300) < 0.05$.  For comparison,
supernova remnants in the published literature commonly show values for
this ratio of $\sim 0.2 - 0.3$.  The lack of \ca2 in the purportedly shocked
regions of NGC~1068 thus presents a puzzle; if the gas has been shocked
it appears that the grains have largely survived intact.  This result
then argues against grain destruction as an important factor in the
generation of strong \fe2 emission in this source.

While the missing \ca2 could be invoked as evidence against the
presence of shocks in the NLR, it may be possible to turn this
argument around and use the survival of grains as an interesting
constraint on the shock physics that is operative in these
environments.  As an example, theoretical studies of dust processing
in shocks indicate that the amount of grain destruction is strongly
influenced by sputtering following betatron acceleration as the grain
traverses the shock (Jones et al. 1994).  The amount of gas
compression and resulting acceleration is influenced by the magnetic
field; stronger $\vec B$ fields parallel to the shock front will
contribute magnetic pressure that reduces the amount of compression in
the shock.  The survival of grains in the NLR may thus lead to useful
constraints on the magnetic field in these environments.

\section{Conclusion}

Emission lines can be used to place useful constraints on the grain
population in Seyfert NLRs.  Features of \fe2 and \ca2 are of
particular interest in this regard, and one point we wish to emphasize
in this contribution is that these two emission tracers are best
analyzed in parallel; more specifically, judgements regarding the role
of grain destruction in enhancing \fe2 emission should also take into
account the constraints provided by \ca2 measurements or limits.

While the \ca2 lines are globally weak in Seyfert nuclei, these features
might be detectable in localized nebular regions where independent evidence
exists for shocks driven by radio plasmas.  Our initial pilot study of
NGC~1068 has produced a null result, however.  This rather surprising
outcome may be employed to place interesting constraints on the nature
of shocks that do, in fact, occur within Seyfert galaxies.

\acknowledgments

Support for this work was provided by NASA and by Ohio University to JS.
We thank Martin Gaskell and the other organizers for a very stimulating
conference.

\clearpage
\noindent
\begin{question}{Martin Gaskell:} Of course it's been known for a long time from
emission-line reddening indicators that there is substantial internal
reddening in NLRs ($E(B-V) \sim 0.5$).
\end{question}
\begin{question}{Ray Norris}  So where do you think the \fe2 is coming from 
in NGC~1068?
\end{question}
\begin{answer}{Joe Shields} I suspect that it's largely a product of a normal,
photoionized, and dusty, NLR; the same conclusion was drawn on more
quantitative grounds by Simpson et al. (1996), who compared measured
line strengths with predictions of a simple Fe~{\sc ii} model atom.
In the near term it should be possible to test this idea more
rigorously, as sophisticated Fe~{\sc ii} models are incorporated into
photoionization codes.
\end{answer}
\end{document}